\documentclass[twocolumn,showpacs,preprintnumbers,amsmath,amssymb]{revtex4}
\usepackage{graphicx}% Include figure files
\usepackage{dcolumn}% Align table columns on decimal point
\usepackage{bm}% bold math
\begin{document}
                                                                                                                             
\preprint{CUPhysics/02/2007}
\title
{A novel approach to study realistic  navigations
on  networks}

\author{Parongama Sen}
\affiliation{
Department of Physics, University of Calcutta,92 Acharya Prafulla Chandra Road,
Calcutta 700009, India.\\
}
%\email{psphy@caluniv.ac.in}

\begin{abstract}
We consider navigation or search schemes on networks which are realistic in the
sense that not all search chains can be completed.
We show that
the quantity  $\mu = \rho/s_d$,
where $s_d$ is the
average dynamic shortest distance   and  $\rho$ the success rate
of completion of a search, is a consistent  measure for the quality
of a search strategy.
Taking the example
of realistic searches on scale-free networks, we find that  $\mu$
scales with the system size $N$  as $N^{-\delta}$,
 where $\delta $
decreases as the searching strategy  is improved.
  This measure is also shown to be
sensitive to
the distintinguishing characteristics of
networks.  In this new approach, a dynamic small world (DSW) effect is said to exist
when $\delta \approx 0$. We show that such a DSW indeed exists in social
networks
in which  the  linking probability is dependent on social distances.
\end{abstract}

\maketitle

\section{Introduction}

During the last few years there has been a lot of activity in the
study of networks \cite{watts,bareview} once it was realised that networks of diverse nature 
exhibit many common features in their underlying structure. 
The most important property that appeared to be commonly 
occurring in such 
networks is the small world property. This means  that if
any two nodes in the network is separated by an average number of $s$ 
steps,
 then $s \propto \ln(N)$, where $N$ is the total number of 
nodes in the network.
In some networks, even slower variation (i.e., sub-logarithmic) scaling has been observed \cite{newman_sub}.

The first indication that networks have small world 
behaviour emerged from an experimental study by
Milgram \cite{milgram}, in which it was shown that 
any two persons (in the USA) can be connected by an
average number of six steps.
Following the tremendous interest in the study
of networks, a new experiment 
has been done to 
verify this property in real social networks \cite{killworth,dodds}.
Some studies which involve  simulations on real networks \cite{adamic_search,hong,geog}  
have been made also.
%  e.g., on a e-mail network using
% possible algorithms etc.   
Parallely, the question of 
navigation on small world networks has been addressed theoretically
in many model networks \cite{klein,adamic1,kim,zhu,moura,watts-search,carmi,thada,clauset}. 

It must be noted that  it is not necessary that
a navigation or searching on a small world 
network would  show the small world property, i.e., the dynamic 
paths  $s_d$ may not scale as  $\ln(N)$.
This is because searching is done using local 
information only while the average shortest distances are calculated
using the global knowledge of the network. 
This was explicitly shown by Kleinberg \cite{klein} in a theoretical study
where nodes where placed on a two dimensional Euclidean space. 
Each node here has connections to its  nearest neighbours  as well as to
neighbours at a distance $l$ with probability  $ P(l) \propto l^{-\alpha}$.
Although the network is globally a small world for a range of values of 
$\alpha$, 
navigation on such networks using greedy algorithm showed a
small world behaviour only at $\alpha = 2$. 
In general the path length showed a sublinear power law increase with $N$.

Recently, the scale-free property of networks has also been found in many
real-world networks which precisely means that the degree distribution 
follows a
behaviour $P(k) \propto k^{-\gamma}$, with the value of $\gamma$ between 2 and 3 for
most networks.
Navigation  on such networks has also been theoretically studied yielding 
the result that
rather than logarithmic, there is a power-law variation of the path lengths 
with $N$
 as in the 
non-scale free Euclidean
network (except for particular algorithms or values of $\gamma$)
\cite{adamic1,kim}. 
%However, for 
%particular values of $\gamma$ and  algorithms (e.g., $\gamma=2$ and an algorithm which involves
%knowledge of second neighbours \cite{adamic1} and the BA network with $\gamma=3$ where the 
%message is   passed to a neighbour with maximum degree \cite{kim}), a logarithmic
%variation can be achieved.
This confirms once again that for navigation in a small world network, 
it is not necessary that the
path lengths will reflect the small world behaviour.

Obviously, a lot depends on the algorithm also, but any  realistic
algorithm is basically local. For Euclidean networks,  
the
greedy algorithm appears to be the most popular one from the findings of 
the  original Milgram experiment as well as that of 
 \cite{killworth,dodds}. Here,  once the source and the target nodes are chosen, the 
strategy is to connect to a neighbour closest to the target. In scale-free
Barab\'asi-Albert (BA) networks \cite{BA}, the two extreme strategies of a random search and highest 
degree search have been used  as well as a preferential search
scheme based on the degree \cite{kim}. Strategies involving other properties 
like   betweenness centrality etc. have also been investigated \cite{thada}.

A relevant question in this context is how to test the quality 
of the search strategy. Noting that the path lengths in all the
studies show a behaviour
$s_d \propto N ^ {\tau_1}$, one may expect that a lower value 
of $\tau_1$ indicates
a better search strategy. 

In realistic searches as in Milgram's experiments and that described in \cite{dodds},
the search may not be successful or complete always.
(In fact for graphs which are not fully connected, there is always a 
finite possibility that a connected path does not exist, even with
global knowledge \cite{latora,holme}.)
Here, apart from the search length, the success rate is also
an important factor. Indeed, most works referring to  
Milgram's 
experiments ignore the fact that very few chains were completed in the
initial experiment (3 out of 60) and better success rate (35\% or more)
was achieved only after
randomness in the search was compromised  
in the sense that the choice of the source persons was made with 
some bias \cite{milgram}.
Many of the theoretical models do not consider the probability of
termination at all and the interest is to find out whatever the search length 
is.

We  argue that for realistic searches, where termination of
search paths is a possibility, one must consider both the success rate as 
well as the path lengths to adjudge the quality of a search strategy.

 We have generated random scale
free networks and used a tunable preferential search strategy which can be extended from 
a random search (RS) to a highest degree search (HDS)  scheme. 
In general we find that $\rho$, the success rate  also follows a power law behaviour such that
$\rho \propto N^{-\tau_2}$. 
On a scale-free network, it is expected that a 
high degree search will improve the search quality.
However, we have shown that a comparison of search strategies based on the behaviour of path lengths alone may lead to an entirely different conclusion.
On the other hand,  we find that the ratio of the success rate to the path length,
  $\mu = \rho/s_d$
has a power law behaviour with $N$, i.e., $\mu \propto N^{-\delta}$ where
$\delta$ decreases as one conducts a higher degree search (section II).
% exponent $\tau_2$ shows a very interesting behaviour
%it may take up both positive and negative values.
We therefore claim that $\mu$ is a reliable measure to test the
quality of a realistic search on a network  involving both the success rate and
the shortest path lengths in an effective manner.

Calculation of $\mu$ on simple networks suggests that $\delta$ can
take up values between zero and one (section III). Corresponding to $\delta=0$ we have the 
best searchability;  this we call a ``Dynamical small world''  (DSW) effect. 
In section IV, we have shown that $\mu$  is indeed 
sensitive to the distinguishing features of networks using the examples 
 of different kinds of scale-free 
networks with the same degree exponent $\gamma$.
Finally,  in section V we have considered a social network in which the nodes possess 
 a ``similarity'' factor and the   
the linking probability of two nodes depends parametrically on it.
Here, we have shown that it is possible to 
obtain  DSW  (i.e.,  $\delta$ assumes values very
close to zero) for a finite range of values of the parameter.

\section{Searching on random scale free networks }

In this section we describe the searching procedure on random scale-free (RSF)
networks, which have no other features other than being scale-free.
Random scale-free  networks are generated 
by   assigning  the degree to each node following 
a scale-free distribution $P(k) \propto k^{-\gamma}$, allowing $k$ to vary from $k_{min}$ to
$k_{max}$. $k_{max}$ is taken to be $N^{1/\gamma}$ while
$k_{min}$ is allowed to vary from 1 to higher values. 
The links are then established randomly between the nodes 
with the given distribution.  Links are assigned in steps; we start with the node 
with the highest degree, in the next step the node with the second highest degree 
and so on. This method has been previously used to generate random scale-free 
networks \cite{lazaros}. 

The search strategy is like this :\\
A source node and a target  node are selected randomly. The source node will 
send the signal to one of its neighbouring nodes provided that node has not
already taken part in the search. This is in tune with Milgram-like experiments. 
\begin{center}
\begin{figure}
\vskip -1cm
\includegraphics[clip,width= 5cm, angle=270]{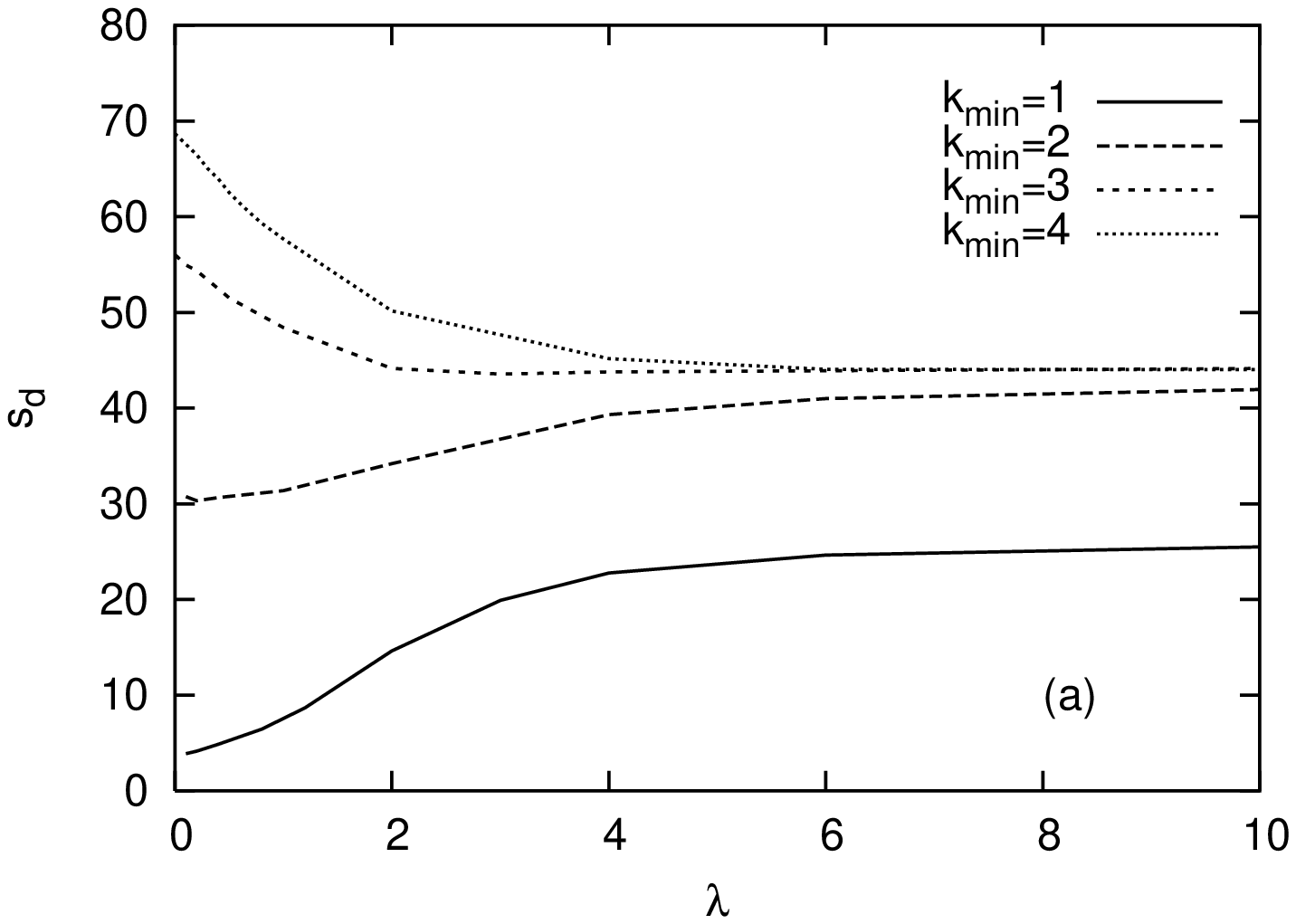}
\includegraphics[clip,width= 5cm, angle=270]{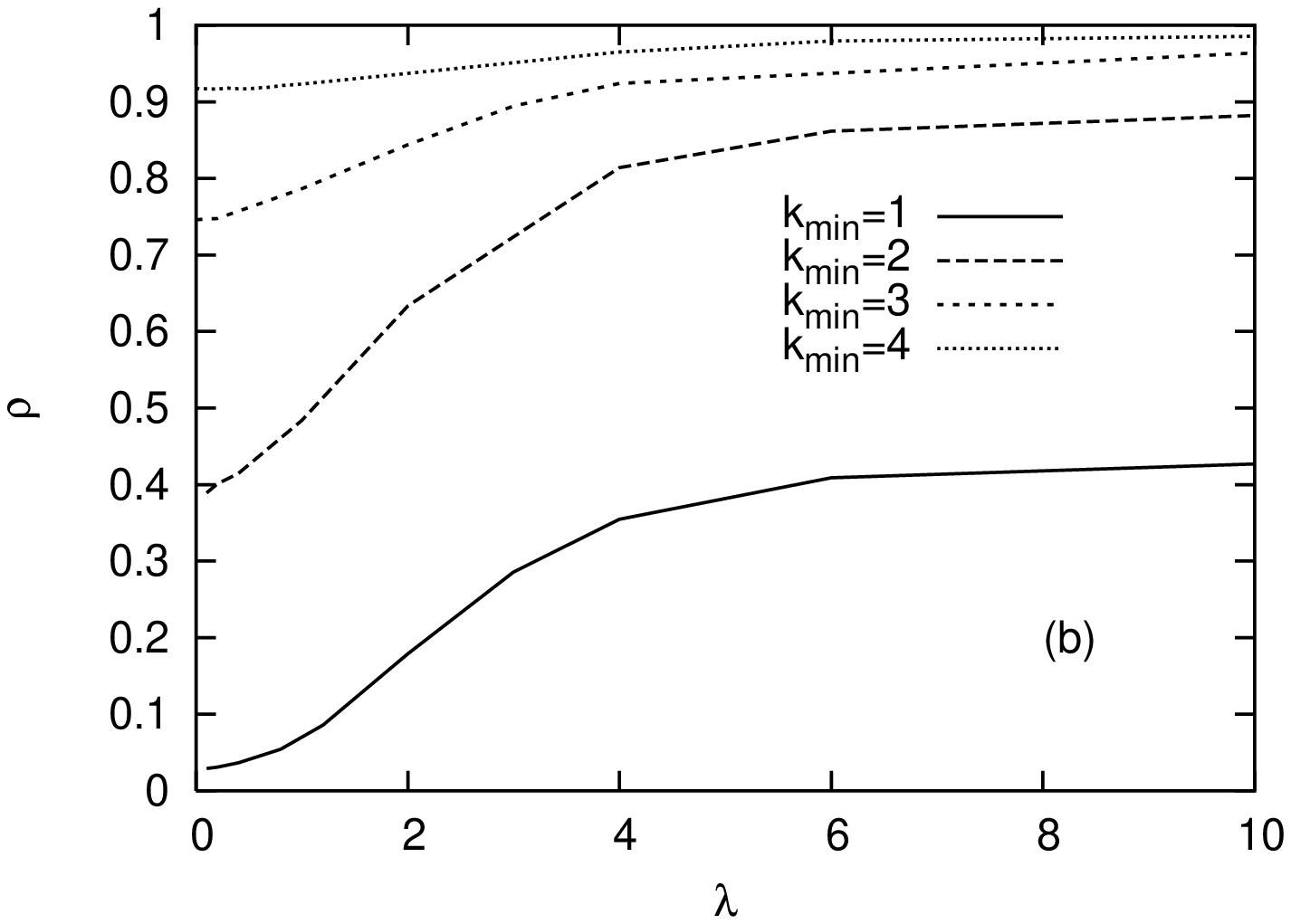}
\caption{ The path lengths (a)  and success rates (b) are shown 
against $\lambda$ for
a random scale-free network for values of $k_{min} = 1,2,3 $ and $4$.
               }
\end{figure}
\end{center}
If one of its neighbour happens to be the target 
itself,  the message will be send to the target. If not, then
the $ith$ neighbour  will  receive the message with a probability  
 $\Pi_i$, where
\begin{equation}
\Pi_i \propto k_i^{\lambda},
\end{equation}
and $k_i$  is the degree of the $i$th node.
This continues and if it happens that the message cannot be passed any more, the chain will 
remain incomplete. 
We take the average dynamic shortest path as the average of all completed chain lengths.
Obviously  $\lambda = 0$ corresponds to a random search while 
$\lambda \to \infty$ gives rise to the highest degree search.
$\lambda = 0,1, \infty$ were considered for the BA network in \cite{kim},
although the algorithm was slightly different.

We first report the results for $s_d$ and $\rho$  for $\gamma= 2$
for a network of $N=1000$ nodes with different values of 
$k_{min}$.
The path length against $\lambda$ shows a very interesting result.
Normally one would expect that the path length will reduce with 
higher degree searches. While this happens for $k_{min} =$ 3 or 4,
for smaller values of $k_{min}$, this is not true. Here, on the contrary,
the path length increases with $\lambda$ (Fig. 1a). For all values of 
$k_{min}$, we  notice that the path lengths
saturate to a constant value  which corresponds to the
highest degree search. This occurs for finite values of $\lambda \sim  8$.
The fraction of successful attempts $\rho$  shows a uniform  behaviour with
$\lambda$; for all values of $k_{min}$, it increases with $\lambda$ (Fig. 1b)
as expected.

The reason why for small $k_{min}$ there is a difference in behaviour of $s_d$ is
quite obvious. When $k_{min}$ is small and a random 
search is carried on, very few chains can be completed,
but if completed, these are essentially  small in length 
as the connectivity on an average is small. On making a higher degree
search, more chains can be completed but at the expense of   increased  
path lengths.

\begin{center}
\begin{figure}[t]
%\vskip -4cm
\includegraphics[clip,width= 6cm, angle=270]{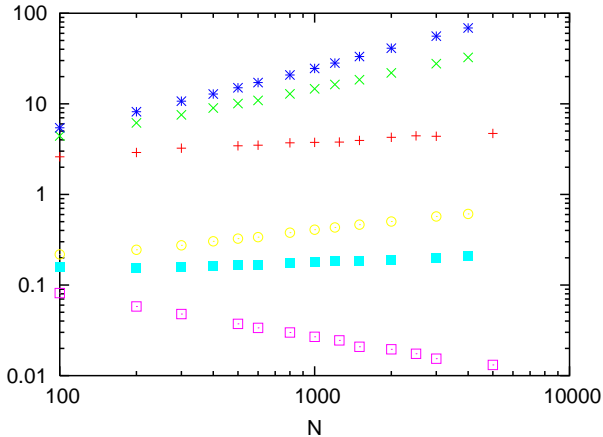}
\caption{The path lengths $s_d$ (set of three curves in the top) 
and the success rates (set of three curves in the bottom) are shown 
against system size 
$N$  for $k_{min} = 1$ for five different values of $\lambda=6.0,~ 2.0$ and $0.0$ (from top to bottom).  }
\end{figure}
\end{center}

The static small world effect is quantatively  measured  by the 
behaviour of the average shortest 
path lengths $s$ (global) with $N$. 
We notice that the dynamic path lengths $s_d$  varies
 with $\lambda$ and in all
cases it shows a  power law increase with $N$ (except for $\lambda=0$ and
$k_{min=1}$ where it has a logarithmoc increase). However, for $k_{min}$ small,
the path length against  $N$ shows that  it varies as $\tau_1$ with 
$\tau_1$ increasing with a higher degree search, which is 
counter intuitive (Fig. 2). So clearly the path length alone is   not a good
 measure of the searchability. The success rate, on the other hand, shows
a change in behaviour  as $\lambda$ is varied.  The exponent $\tau_2$ shows a 
decrease
with $\lambda$; for small values of $\lambda$, $\tau_2$ 
  is positive and it goes on to assume negative values 
for higher $\lambda$.

However, when we 
study the behaviour of the ratio $\mu = \rho/s_d$ as a function of  $N$, 
we find a consistent behaviour  $\mu \propto  N^{-\delta}$, where 
$\delta $ decreases as $\lambda $ is increased (Fig. 3a).

For higher values of $k_{min}$, 
path length and success rates show expected behaviour with $N$; both 
improve with $\lambda$.
$\delta$ also  decreases with $\lambda$ monotonically 
as in the case of $k_{min} =1$ (Fig 3b). 
Hence one can regard $\mu$  as a  measure to adjudge the
quality of the search strategy in general.
For higher values of $k_{min}$ (which implies higher connectivity),
variation of $\delta$ with $\lambda$ is not that appreciable showing that
for networks with larger connectivity, the sensitivity 
to degree-based  algorithm is not remarkable.  
 
\begin{center}
\begin{figure}
\includegraphics[clip,width= 5cm, angle=270]{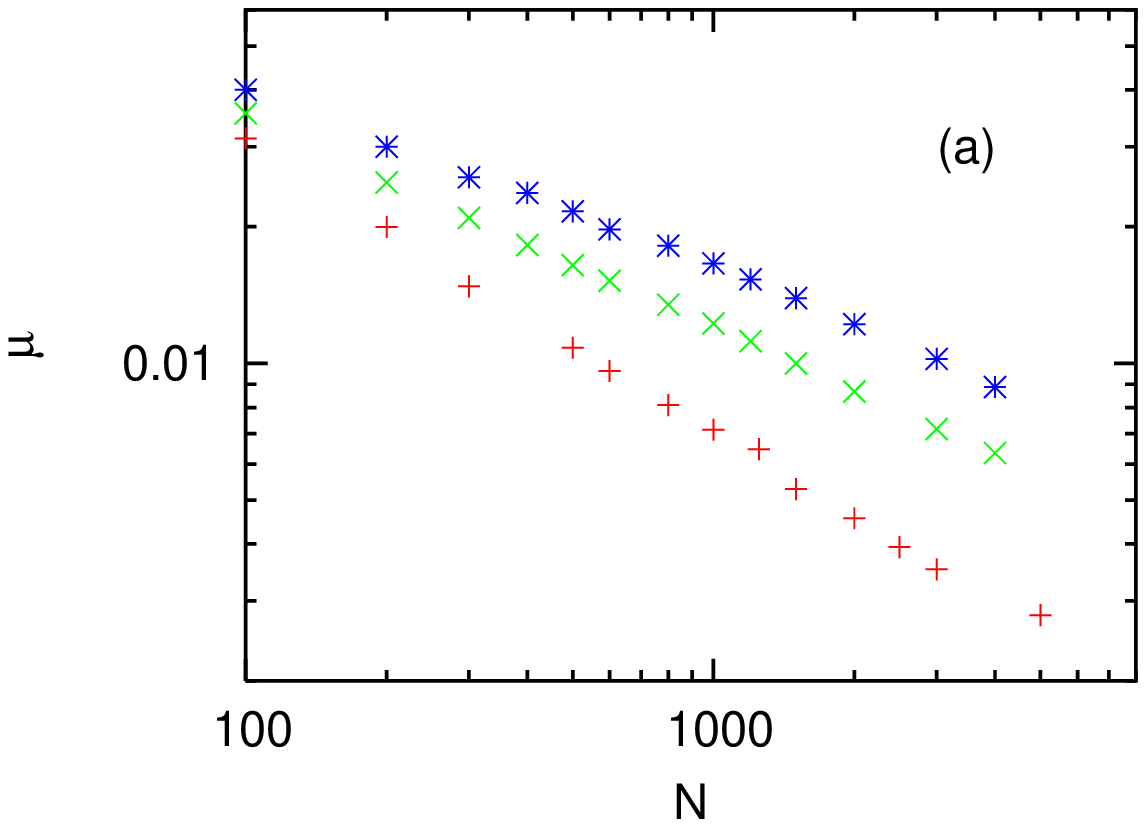}
\includegraphics[clip,width= 5cm, angle=270]{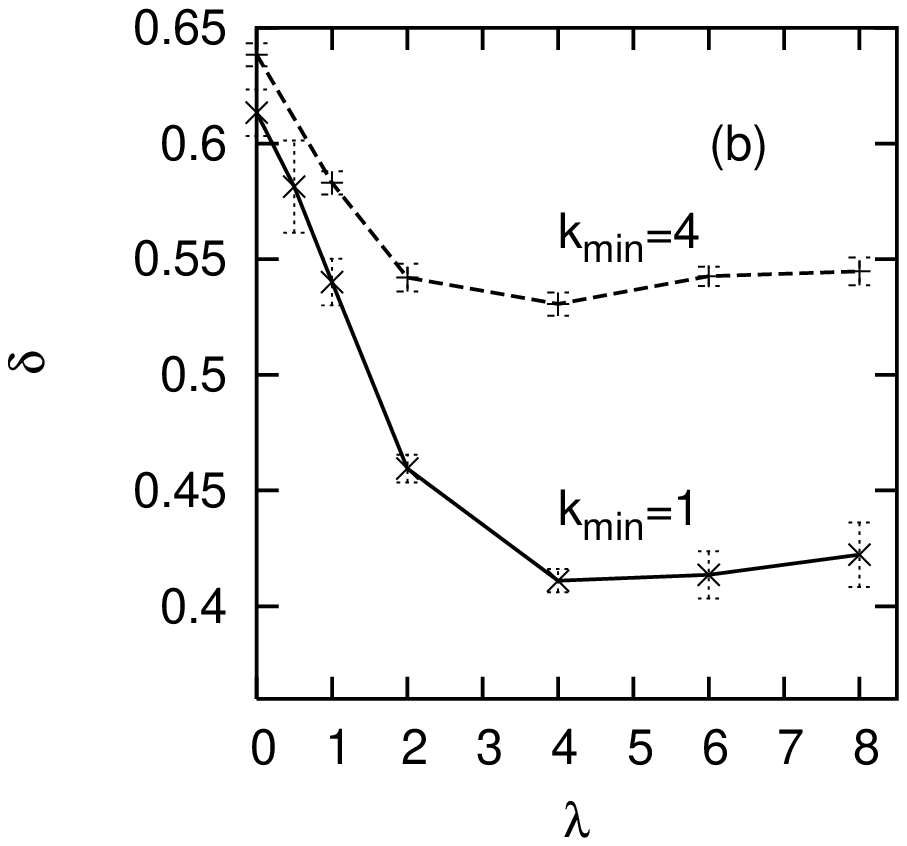}
\caption{(a) The ratio $\mu = \rho/s_d$ against $N$  shown for $\lambda$ = 6.0,  2.0  and 0.0 (from top to bottom) for a RSF network 
with $k_{min} = 1$. In (b) The corresponding exponents $\delta$ are shown   
against $\lambda$ for $k_{min} = 1$ and
$4$. For both cases, there is a monotonic decrease of $\delta$ with 
$\lambda$.
               }
\end{figure}
\end{center}

\section{Calculation of $\mu$ and $\delta$ in some simple cases}

The searchability factor $\mu$ and hence $\delta$  can be calculated in some simple cases.
Although the examples may seem too idealised, the results
help us to understand the bounds of $\delta$. 

1. Search on a regular network  with nearest neighbour interactions only.
The simplest case is a one dimensional chain (Fig. 4a).  
On any Euclidean network,
a greedy allgorithm,  i.e., passing the message to a 
node closest to the target is meaningful. Here as the links are between
nearest neighbours only, the  search path length
is  equal to $l$ where $l$ is the distance separating them. In general $l = pN$ where
$0 \leq  p \leq 1$. The success rate is hundred percent here such that
\begin{equation}
\mu_{reg} = 1/pN 
\end{equation}
and $\delta = 1$ in this case.

Note that in this case, since the message cannot be passed to the same node more
than once, the search  must take a unique path, it has to be directed 
towards the target in steps of 1 (shown by the arrows in Fig 4a).
Thus random search is not conceivable here.
If the  message can be passed to the same node
more than once, 
 one can use a random search and the situation is analogous to a random walker. The
path length will now be proportional to $l^2$ and the corresponding $\delta = 2$.
This is in fact the worst case scenario: however, it does not belong to 
the class of navigation considered here where messages can be received only once.

\begin{center}
\begin{figure}
\includegraphics[clip,width= 7cm]{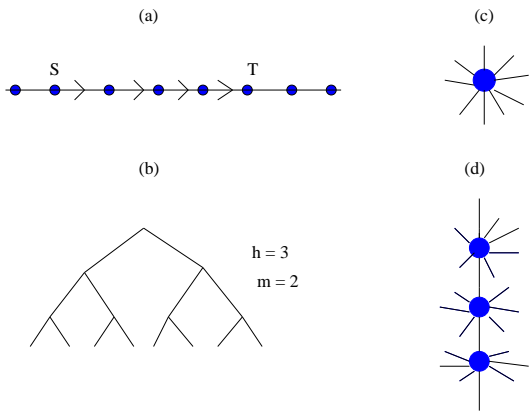}
\caption{Simple networks for which $\mu$ can be calculated.
In (a), a regular network in one dimension is shown, $S$ and $T$ are the source and target nodes respectively. The unique path from $S$ to $T$ 
is shown. In (b), a tree network with number of levels $h$ = 3 and
branching ratio $m=2$ is shown. In (c) a gel or star network and in (d), 
a chain of such gels (or hubs) have been shown.}
\end{figure}
\end{center}

2. Search on a tree with uniform branching ratio: Let us imagine a tree with
branching ratio $m$ (Fig. 4b) and with $h$ levels. Let us consider the case of sending a 
message from the top (zeroeth level) to a node at the bottom 
($h$th) level. 
The probability that a message is sent to this  node at the $h$th  level from 
the top is $(1/m)^h$. Noting that the number of nodes $N\approx m^{h+1}$, 
\begin{equation}
\mu_{tree} \approx N^{-1}\ln(m)/\ln(N), 
\end{equation}   
implying $\delta = 1$ here.
More generalised cases  where search from arbitrary sources and targets are
considered may give different values of $\delta$. For example, if one considers
search paths of length 1 only, then the success rate is $\sim 1/m$ 
giving $\delta=0$. Larger paths will have a different $\delta$  and $\delta$  in principle can vary from 0 to 2.
If one takes all the possibilities, i.e., all possible pairs of source and targets, then the average $\delta$ can be 
estimated from the following expression:
\begin{equation}
\langle \mu \rangle = \frac{\sum_{n=1}^{2h} \mu_n f_n}{\sum_{n=1}^{2h} f_n}
\end{equation}
where $\mu_n$ is the ratio $\rho/s_d$ for two nodes separated by $n$ edges and $f_n$ the number
of such pairs. The terms in the numerator decrease with $n$;
for $h >> 1$, $f_1 \approx m^{h+1}$, $\mu_1 = 1/m$, while $f_{2h} \approx m^{2h}$ and
$\mu_{2h} = \frac{1}{2hm^{2h}}$. Thus the leading order term in the numerator is $O(N)$  
while the leading order term in the denominator is $O(N^2)$ giving $\delta = 1$
for very large trees.
%Taking all the possibilities, for very large trees (i.e.,  $h>>1$), 
%t can be shown that on an  average $\mu \sim N^{-1}$ to the leading order
%n $N$ giving $\delta =1$.  

3. Search on a gel - Here we consider a network in the form of a gel or a star (Fig. 4c)
where there is a central node to which all other nodes (having only one link)
are attached. Again path lengths are of the order 1 and success is
a surity. Hence $\delta  = 0$ here.

4. A chain of hubs or gels: 

We consider a idealised chain of hubs or gels, it contains  $l$ hubs 
with degree 
$k$ each (Fig. 4d); the total number  of nodes $N = kl$. 
Now a search path length will be of the order of $l$, say $\alpha l$.
If we consider a 
random search, 
the probability of completing the search is $(1/k)^{\alpha l}$.
Or, 
\begin{equation}
\mu_{hubs} = \frac{1}{k}^{\alpha N/k}/\alpha \frac{N}{k}.
\end{equation}
The thermodynamic limit $N \to \infty$ may correspond to two different
cases.  The first is when 
 $l$ is small but $N/k$ is finite.
The search probability is small (but non-zero)  and $\mu$ 
 does not vary with $N$  (as the ration $N/k$ remains finite)
giving $\delta = 0$. 
 
On the other hand if $l$ is  large and $k$  finite,   
$\ln(\mu) \approx \alpha N/k \ln (k)$ such that
$\mu$ goes exponentially to zero rather than as a  power law.

 The above discussion shows that when $\mu \sim N^{-\delta}$, 
the value of $\delta$ for 
small world networks lies between $0$ and $1$  when the 
search is done randomly. Obviously, for a more efficient algorithm $\delta$ 
is lesser than 
that for the random search and thus  it can be said that for searches in which repetition
of messengers is not allowed, $0 \leq \delta \leq 1$. A dynamic small world 
in this perspective is one in which $\delta = 0$.

In this section,  the simple networks which have been considered have no loops and
hence the paths from one node to another is unique. Thus
the important quantity determining $\mu$ is essentially the success rate.
In the  trivial cases like the linear chain and the gel, 
even the success rate is deterministic. 
 On 
more complicated networks, both $s_d$ and $\rho$ will have to be computed
probabilistically.

\section{Sensitivity of $\mu$ to the distinguishing 
characteristics of model networks}

In this section we show that our measure of searchability $\mu$ is sensitive to the
distinguishing features of different networks. We have considered networks 
which are equivalent as far as the 
degree distribution is concerened in the sense they are all scale-free 
with the same exponent but differ in certain other aspects. 
We have  selected scale-free
networks with the same exponent $\gamma=3$ and used a HDS algorithm
which is meaningful in a scale-free network.
In the  two subsection we give a brief desciption of the models we have 
considered and the results obtained.

\subsection{Models}

We have considered three different types of scale free models:

1. The Barab\'asi Albert (BA) network: 

The BA network is grown using a preferential attachment scheme. Here,
we have started with a single node and nodes are attached to the network
one by one. An incoming node gets attached to the $i$th existing node
with the probability
\begin{equation}
{\cal{P}}_{BA} \propto k_i,
\end{equation}
where $k_i$ is the degree of the $i$th node at that time.
This network has a
has a tree structure as we allow only one link to an incoming node.

2. The random scale free (RSF) network:

This network is the same as that described in section II.
We have taken $\gamma=3$ here and kept $k_{min} = 1$
such that it is comparable to the BA network.

%The other two are the BA network  (which has a 
%history of being a grown
%network with preferential attachment)

3. Scale free assortative network (SFA)

The third network which we have considered is a scale-free model with 
tunable assortativity \cite{newman_asso}.
This scale-free  assortative (SFA) 
network is generated by modifying slightly the 
method of generating the RSF  network.    
  To each node, the degree  is  assigned from a scale-free distribution 
as in RSF. Let $k_i$ and $k_j$ be the degrees assigned to nodes $i$ and $j$. 
While establishing the links of the $i$th node, 
attachment to the $jth$ node  will be definitely  made 
if $k_i=k_j$.  Otherwise they will be linked  
 with a probability 
\begin{equation}
{\cal{P}}_{SFA}  \propto |k_i-k_j|^{-\sigma}.
\end{equation}
Here $\sigma$ controls the assortativity factor. For $\sigma =0$, there
is no assortativity and for $\sigma > 0$ $(<0)$, the
network is assortative (disassortative). 

It may be noted that both the RSF and the BA networks have zero assortativity
\cite{newman_asso}.

\subsection{Results}

We employ the HDS algorithm, i.e, a node always passes on the information to
the neighbour with the highest degree. In the BA network, 
for  system sizes considered here,  
$s_d$ 
shows a power law increase with $N$ with an exponent $\tau_1 <  0.1$ 
which becomes smaller 
with $N$ showing that it approaches a logarithmic increase with $N$ for 
large $N$. This is to be expected as it has a tree structure.  
 (In \cite{kim} it was shown that even with loops, the 
highest degree search scheme in the BA network results in
a logarithmic increase of the path lengths with $N$.) 
$\rho$ follows
a 
power law decay with $N$ such that  $\mu \propto N ^{-0.6}$.  

%In order to have a comparable random scale free network, we use a $k_{min}$ value
%equal to 1 and $\gamma=3$. 
For the RSF network, we again apply HDS and  find that 
 in this case  $\mu \propto N ^{-0.93}$, although the
path length varies logarithmically with $N$.
(It
may be mentioned here that for the RSF, 
as $\gamma$ is increased from 2 to 3, the
searchability becomes less efficient in terms of $\mu$ and
also becomes less sensitive to the factor $\lambda$.) 

Thus in terms of simply the path lengths the RSF  and BA networks are
equally searchable  which
is  counterintuitive. On the other hand $\mu$ gives a more realistic
result.

In the third network, even when it is weakly assortative ($\sigma=0.1$), we find that $\delta$ shows a
deviation from the random network value.
Here $\delta \simeq 0.87$ which means that it is more searchable than the 
RSF but less than the BA network with the same algorithm.
The behaviour of $\mu$ with $N$ for different networks is shown in Fig. 5.
\begin{center}
\begin{figure}[t]
\includegraphics[clip,width= 5cm, angle=270]{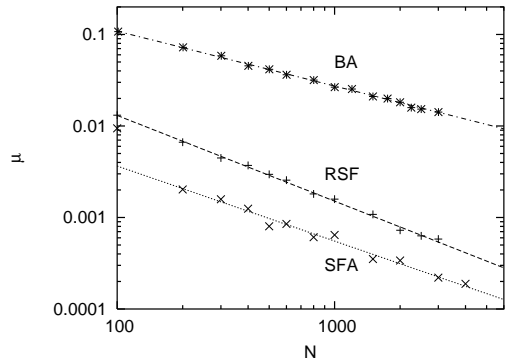}
\caption{The behaviour of $\mu$ against $N$ for three different
scale-free networks (BA - Barab\'asi-Albert, RSF - Random scale-free and 
SFA - Scale free assortative) with identical degree distribution $P(k) \propto k^{-\gamma}$ ($\gamma=3$) is shown. The straight lines correspond 
to the best fits.} 
\end{figure}
\end{center}

As already mentioned, in the BA as well as in the RSF networks,
the path length shows a logarithmic increase with $N$ but 
since $\delta$ is appreciably greater than zero, 
neither of these two  networks is  an ideal dynamic small world.
The random scale-free network  
has a worse searchability because of the sharper decay of the success rate.
In the SFA  network, there is a tendency of clustering of high degree 
nodes thereby allowing loops. Hence the path length does not scale logarithmically with $N$.
However, one only needs to know the behaviour of $\mu$ to comment on 
the searchability and
here it shows a slower decay with $N$ compared to the RSF network although the magnitude of $\mu$ is lesser.

\section{Dynamic small world effect in a  model social network }
 
Till now we have discussed search strategies on network
which are basically characterised by their degree distribution.
Social networks, which are found to be scale-free in many
cases, have some additional features. It has been found that
in a society, people have different characteristics and
the links depend on these bonds. These characteristics can be
related to profession, hobbies, geographical locations etc.

We have considered a very simplified picture with one such characteristic
which we call the similarity factor $\xi$ of the individuals varying between 0 and 1 randomly. In a scale-free network with given distribution $P(k)=k^{-\gamma}$, we
first assign the degrees as well as the similarity factors to the nodes.
The bonding between two nodes,
subject to this distribution is now made according to
\begin{equation}
{\cal{P}}_{i,j} \propto |\xi_i -\xi_j|^{-\alpha}.
\end{equation}

Thus we will have a scale-free network of degree exponent $\gamma$
in which similar nodes will try to link up for $\alpha > 0$.
The similarity factors $\xi_i$ can be looked upon as coordinates in
a one dimensional lattice in which case the network is  scale-free network as
well as Euclidean.

In the earlier cases only  degree based strategies have been considered
as the networks did not have any other characteristic feature.
But as shown in \cite{dodds} such a strategy is  rather
artificial for social networks. Here we have considered two different strategies;
one is the standard HDS while the other one is a greedy algorithm based on the
similarity factor.
In the latter,  a node sends the signal to a neighbour which is  most similar to the destination node.
We call this the highest similarity search (HSS).
We investigate how the searching strategies are dependent on $\alpha$ based
on the value of $\delta$.

We consider networks with $\gamma=2$ and $k_{min} = 2$ here and
vary $\alpha$ to obtain networks  varying on the basis of similarity based 
bondage 
between the nodes for the same degree exponent. We have shown in section IV
that the measure $\mu$ is able to distinguish  between
networks of same degree exponent  which differ in other characteristics and
hence it is meaningful to employ this measure to investigate  
the searchability based on different algorithms here.
\begin{center}
\begin{figure}
\includegraphics[clip,width= 5cm, angle=270]{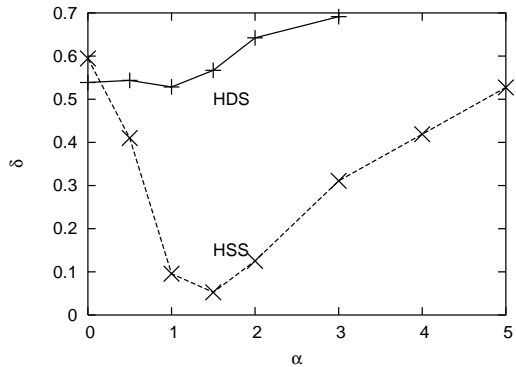}
\caption{The values of $\delta$ against $\alpha$ shows 
a minimum where the network is most searchable for the HSS. For the HDS,
the values of $\delta$ are generally higher and also less sensitive to $\alpha$.}
\end{figure}
\end{center}

Again we find that $\mu$ shows a power law decay with $N$ for both strategies in general. However, for the highest degree search, the value
of $\delta$ shows marginal dependence on $\alpha$ and the
network is moderately searchable with this strategy as 
$\delta$ lies between 0.5 and 0.6. Using the HSS on the other hand we find
an appreciable variation of $\delta$ with $\alpha$. In fact, we find that
for $ 1 < \alpha < 2$, the network is highly searchable  as $\delta$ is $O(0.1)$
here. For $\alpha = 1.5$, $\delta$  becomes 
lesser  for higher $N$ values indicating that it might go to zero 
in the thermodynamic limit giving rise to a truly dynamical small world effect.
 
It is true that not all social networks are scale-free. However,  our 
studies 
lead to the important result that 
{\it {even for scale-free networks}}, if there is a social
distance dependent linking scheme, the degree based search is rather
inefficient (compared to the one based on the social characteristic).
Thus we conclude that for any social network, the degree based search will
be less useful. 

Here we have considered networks of size $N \leq 10000$ with a cutoff in the
degree distribution at $N ^{1/\gamma}$ ($\gamma=2$). In order to simulate
the friendship or acquintance networks, where in principle, the whole population
of the earth is involved, it is better to use a degree distribution
with an exponential cutoff or simply exponential decay to make
the study more realistic \cite{amaral}.

\section{Summary and Conclusions}

In this work we have proposed a new method to study the
searchability on networks where the search may not be complete always
as transmission of signals cannot be repeated to the same node.
In real searches, failure to complete a chain may be due to various 
 other
reasons \cite{dodds}.
The motivation comes from Milgram-like experiments of searches where
the success rate is really low when the source nodes are randomly selected. 
We first show that only the
study of search path lengths as a function of the system size is not
enough to determine the searchability of a network.
 Rather, the ratio of the success rate to the
path length, $\mu$,
is a more
reasonable measure. We have established this by executing
searches on scale-free networks but the significance of $\mu$ is
by no means limited to scale-free networks only.

We have shown that in general $\mu$ behaves as $N^{-\delta}$
with $\delta$ showing a decrease for a better algorithm.
The value of $\delta$ is indicated to be between 0 and 1.
The zero value signifies what we have termed a dynamic small world effect.

Extending the study to  networks in which social distances matter, 
we have shown that
algorithms based on social distances rather than on degree
tend to be more successful in general. 
This is again consistent with the findings of \cite{dodds}.
Here we have introduced a parameter $\alpha$ which governs
the similarity dependent linking probability between two nodes.
We find that  $\mu$, 
the searchability, is maximum at  $\alpha \approx 1.5$, where $\delta$ 
is very close to zero. This corresponds to a network where the 
similarity factor controls the linkings to a moderate extent. This
seems realistic; in a society where only very similar people 
share links (i.e., $\alpha$ has very large values) the searchability cannot be 
high. On the other hand, for low values of $\alpha$, the network 
is almost random, where the searchability is again not very high.

Our studies are motivated by the finding of \cite{milgram,killworth,dodds} and \cite{geog}.
We have obtained reasonable  qualitative agreement  
with the experimental observations (specifically with that of \cite{dodds})
  as already mentioned. 
Further comparison is not feasible for various reasons. In real experiments, 
several
reasons for termination of the search exist giving rise to a 
lower success rate.  
Also, the main emphasis of the present paper is the scaling behaviour 
of the relevant quantities 
with the system size $N$, which has not been studied in these experiments.

Acknowledgement: Financial support from CSIR grant no. 3(1029)/05-EMR-II
  is acknowledged.
The author thanks  K. Bharadwaj and K. Basu Hajra for discussions. 

\end{document}